\documentstyle[preprint,aps]{revtex}

\begin{document}

\draft
\preprint{CERN-TH.6860/93}

\title{Successive Combination Jet Algorithm\\
For Hadron Collisions
}
\author{Stephen D.~Ellis\cite{byline}\\
}
\address{
Theoretical Physics Division, CERN\\
CH - 1211 Geneva 23 \\
SWITZERLAND
}
\author{Davison E.~Soper}
\address{
Institute of Theoretical Science\\
University of Oregon \\
Eugene, OR  97403, USA
}
\date{\today}
\maketitle
\begin{abstract}
Jet finding algorithms, as they are used in $e^+ e^-$ and hadron
collisions, are reviewed and compared.  It is suggested that a successive
combination style algorithm, similar to that used in $e^+ e^-$ physics,
might be useful also in hadron collisions, where cone style algorithms
have been used previously.
\end{abstract}

\narrowtext

\section{Introduction}

The measurement of jet cross sections has provided useful tests of Quantum
Chromodynamics both at hadron colliders and at electron-positron colliders.
The observed jets provide a view of the underlying hard quark and gluon
interactions that occur at very small distance scales. However, this view
is inevitably clouded by the subsequent long distance showering and
eventual hadronization of the primary quarks and gluons. Furthermore, since
the quarks and gluons carry non-zero color charges and the final hadrons do
not, there can be no {\em unique} association of a jet of hadrons with a
single initial quark or gluon. Nevertheless, with a suitable definition of
the jet cross section one hopes to minimize the effect of long distance
physics and of the inherent jet ambiguities and obtain a fairly precise
picture of the short distance dynamics.

Although the basic hard scattering processes studied in hadron-hadron and in
electron-positron collisions are much the same, the overall event structure
is quite different. In the $e^{+}e^{-}$ case the initial state is purely
electromagnetic and the entire final state can be thought of as arising from
the short distance interaction of the virtual photon with the quarks. In
this sense all of the hadrons in the final state are associated with the
hard scattering process. In contrast the overall structure of the
hadron-hadron case is much more complex. Of the large number of initial
state partons, only one ``active parton'' from each incident hadron
participates in the hard scattering process. Thus only a fraction of the
hadrons in the final state are to be (loosely) associated with the hard
scattering process, with the remainder ascribed to the ``underlying event.''
This second contribution corresponds to the soft interactions of the
remaining partons in the incident hadrons and, in first approximation, can
be treated as uncorrelated with the hard process. The active partons also
produce additional radiation in the form of initial state bremsstrahlung
that is not present in $e^{+}e^{-}$ events. The underlying event plus the
initial state radiation produce the characteristic ``beam jets'' of hadron
collisions: particles with small momenta transverse to the beam axis, but
possibly large momenta along the beam axis. The long distance soft
interactions responsible for the observed color singlet hadrons will, of
course, result in some degree of dynamical coupling between all of these
components. There will also be an essentially kinematical correlation
induced by the fact that the jet selection or trigger process will generally
be biased to choose events where the beam jets have higher than average
global $E_T$ and multiplicity ({i.e.}, the underlying event is noisier
than average).

These differences between the event structure of $e^{+}e^{-}$ collisions and
hadron-hadron collisions have, quite naturally, led to differences in the
way jet definitions have been employed in the two cases. One might
categorize the differences as follows.

First, the cross sections studied are different. In $e^{+}e^{-}$ collisions,
where the entire event arises from an initially small number of energetic
partons, one typically works with {\em exclusive} jet cross sections
describing the production of exactly $n$ jets and nothing else. In
hadron-hadron collisions the practice has been to measure {\em inclusive}
large $p_T$ jet cross sections, that is, cross sections to make $n$ jets
with specified properties plus any number of other unobserved jets or
particles not in jets.

Second, the variables used are different. For $e^{+}e^{-}$ annihilation, one
wants to emphasize rotational invariance. Thus the natural variables are
energies $E$ and polar angles $\theta ,\phi $. For hadron-hadron collisions,
one wants to emphasize invariance under boosts along the beam axis since, in
fact, the c.m.\ frame of the hard scattering is typically moving in the
hadron-hadron c.m.\ frame. Thus the natural variables are transverse momenta
$p_T$ or the corresponding ``transverse energy'' $E_T\equiv E\sin \theta $,
azimuthal angle $\phi $ and pseudo-rapidity $\eta =-\ln (\tan (\theta /2))$.

Third, the jet definitions or algorithms employed to precisely define the
(otherwise ambiguous) jets tend to be correspondingly different. As implied
above in $e^{+}e^{-}$ collisions one normally uses a jet definition that
associates {\em every} final-state hadron {\em uniquely} with one of the
jets. In hadron-hadron collisions producing high $p_T$ jets, only a small
fraction of the final state hadrons are associated with the high $p_T$ jets.
The other particles present in the event can be thought of as associated
with the ``beam jets'' introduced earlier. One wants to keep the high $p_T$
jets distinct from the hadronic debris in the beam jets. For this reason,
one has typically used a cone definition \cite{snowmass}, which was, in
fact, inspired by the original theoretical definition for jets in $e^{+}e^{-}
$ collisions \cite{StermanWeinberg}. A jet in this definition is a set of
particles whose momentum vectors lie within a certain angular cone. Such a
definition suppresses the effect of the beam jets, since only a small
fraction of the low $p_T$ particles in the beam jets will fall into the cone
of a high $p_T$ jet. Furthermore, since this contribution is essentially
determined by geometry, it is relatively easy to estimate. In the case of $%
e^{+}e^{-}$ collisions, the jet algorithm typically used experimentally is
rather of the ``successive combination'' variety first introduced by the
Jade group at DESY \cite{Jade}. In this kind of definition, one recursively
groups sets of particles with ``nearby'' momenta, as defined by some
measure, into larger sets of particles. The initial sets consist of just one
particle each. The final sets are the jets.

In this paper, we discuss a jet definition for hadron collisions that makes
use of some of the ideas used in jet definitions in $e^{+}e^{-}$ collisions.
The definition is essentially that proposed by S.\ Catani, Yu.\ Dokshitzer,
M.\ Seymour, and B.\ R. Webber \cite{Webber}, adapted for the measurement of
inclusive, rather than exclusive, jet cross sections. We first define the
algorithm. Then we comment on some of its features. Finally, we provide some
evidence that this definition may have advantages compared to the cone
definition that is currently the standard for hadron collisions.

\section{The algorithm}

We consider hadron collisions in the hadron-hadron c.m. frame with the
z-axis taken in the beam direction. We represent the final state of the
collision as consisting of a starting set of ``protojets'' $i$ with momenta $%
p_i^\mu$. The starting $p_i^\mu$ may be the momenta of individual particles,
or each $p_i^\mu$ may be the total momentum of the particles whose paths are
contained in a small cell of solid angle about the interaction point, as
recorded in individual towers of a hadron calorimeter. In either case, we
have in mind that the masses $[p_i^\mu p_{i\mu}]^{1/2}$ are small compared
to the transverse momenta $p_{i,T}$, so that the $p_i^\mu$ are essentially
lightlike. Each protojet is characterized by its azimuthal angle $\phi_i$,
its pseudorapidity $\eta_i = -\ln(\tan(\theta_i/2))$, and its transverse
energy $E_{T,i}= |\vec p_{T,i}|$.

Starting with the initial list of protojets, the jet algorithm recursively
groups pairs of protojets together to form new protojets. The idea is that
protojets with nearly parallel momenta should be joined, so that they will
eventually form part of the same jet. The algorithm also determines when,
for a particular protojet, joining should cease. This protojet is then
labeled a completed ``jet'' and is not manipulated further.

The algorithm depends on a parameter $R$, which should be chosen to be of
order 1. This parameter is analogous to the cone size parameter in the cone
algorithm.

The algorithm begins with a list of protojets as described above and an
empty list of completed jets. It then proceeds recursively as follows:

\begin{enumerate}
\item For each protojet, define
\begin{equation}
d_i = E^2_{T,i}
\end{equation}
and for each pair of protojets define
\begin{equation}
\label{dij}
d_{ij} = \min(E^2_{T,i},E^2_{T,j})\ [(\eta_i-\eta_j)^2 + (\phi_i
- \phi_j)^2]/R^2.
\end{equation}

\item Find the smallest of all the $d_i$ and
$d_{ij}$ and label it $d_{{\rm min}}$.

\item If $d_{{\rm min}}$ is a $d_{ij}$, merge protojets
$i$ and $j$ into a new protojet $k$ with:
\begin{equation}
E_{T,k} = E_{T,i}+E_{T,j} \label{ET1}
\end{equation}
and
\begin{mathletters}
\label{axis1}
\begin{eqnarray}
\eta_{k}&=& [E_{T,i}\eta_{i}+E_{T,j}\eta_{j}]/E_{T,k},
\\
\phi_{k}&=& [E_{T,i}\phi_{i}+E_{T,j}\phi_{j}]/E_{T,k}.
\end{eqnarray}
\end{mathletters}
\item If $d_{{\rm min}}$ is a $d_i$, the corresponding
protojet $i$ is ``not mergable.'' Remove it from the list of protojets and
add it to the list of jets.

\item Go to step 1.

\end{enumerate}

The procedure continues until there are no more protojets. As it proceeds,
it produces a list of jets with successively larger values of $d_i =
E_{T,i}^2$.

\section{Comments}

The algorithm above produces a list containing many jets for each event.
However, only the jets with large values of $E_T$ (which are the last to be
added to the jet list) are of much physical interest. The jets with smaller $%
E_T$ are ``minijets'' or just random debris from the beam jets. This
situation is fine for the construction of an inclusive jet cross section.
Consider, for instance, the one-jet inclusive cross section $d\sigma /dE_T$
for, say, $E_T=100$ GeV at $\sqrt{s}=1800$ GeV. We first note
that the high value of $E_T$ tells us that this jet is a signal of a short
distance process. Second, we recall that in hadron collisions the
probability for a parton collision decreases very quickly as the value of
the parton-parton c.m.\ energy $\sqrt{\hat s}$ increases. Thus it is very
unlikely that an event with a 100 GeV jet contains other high $E_T$ jets
beyond a second jet with $E_T\approx 100$ GeV that is needed to
balance the transverse momentum. That is to say, the one jet inclusive cross
section is primarily sensitive to the highest $E_T$ jets in hard scattering
events, even though the jet list for each event contains many jets covering
a wide range of transverse energies.

It is crucial that a jet cross section be ``infrared safe.'' That is, when
the cross section is calculated in QCD at the parton level, the cross
section must be finite, despite the infrared divergences present in the
Feynman diagrams. At the level of the physical hadrons, this means that the
cross section is not sensitive to long distance effects. The infrared
divergences in Feynman diagrams come from configurations in which a parton
emits a soft gluon, with $q^\mu \to 0$, or in which an outgoing parton
divides into two collinear partons, or in which an incoming parton emits
another parton that carries away a fraction of its longitudinal momentum but
no transverse momentum. The probability for one of these configurations to
occur is infrared sensitive, and infinite in fixed order
perturbation theory. However,
unitarity dictates that the sum of the probabilities for one of these
configurations to happen or not to happen is 1. For this reason, infrared
safety is achieved if the measured jet variables do not change when an $%
E_T\to 0$ parton is emitted or when a parton divides into collinear partons.
We note that $d\sigma /dE_T$ and similar jet cross sections produced using
the algorithm described in the previous section will have this property. If
a parton divides into two partons with collinear momenta, then the algorithm
immediately recombines them, producing the same result as if the parton had
not divided. Similarly, an $E_T\to 0$ parton may wind up in one of the high $%
E_T$ jets or it may be left by itself, but in the limit that its $E_T$ tends
to zero it does not affect the transverse energy or angle of the high $E_T$
jet.

We have discussed the one jet inclusive cross section. For the two jet
inclusive cross section, it would be sensible to pick the two jets in each
event that have the highest transverse energies, just as has been done for
jets defined with the cone algorithm. By analogy, in defining the one jet
inclusive cross section one might be tempted to pick only the most energetic
jet in each event. However, at the Born level there are two jets with
exactly equal transverse energies. Which one winds up with the most
transverse energy is affected by a long-distance process, the emission of
very low $E_T$ gluons. Thus the resulting cross section would not be
infrared safe.

The algorithm defined above is very similar to the simplest version of the
various options discussed by Catani {et al.} \cite{Webber}, which are
generalizations of the ``Durham'' algorithm \cite{Durham} for $e^{+}e^{-}$
annihilation. The differences arise primarily from questions of emphasis.
Catani {et al.} take the approach that the analysis should be kept as
similar as possible to earlier $e^{+}e^{-}$ work. Thus they fix the
parameter $R$ at the value 1 and focus on {\em exclusive} jet cross
sections. They also introduce two further parameters. The first additional
parameter is $d_{{\rm cut}}$. When the smallest $d_i$ or $d_{ij}$ is larger
than $d_{{\rm cut}}$, their recursion halts. The jets that have been
generated thus far, all of which have $E_{T,i}^2<d_{{\rm cut}}$, are
regarded as part of the beam jet. The remaining protojets, all of which have
$E_{T,i}^2>d_{{\rm cut}}$, are treated as resulting from the hard scattering
process. These protojets are then resolved into the final ``jets'' in direct
analogy to the $e^{+}e^{-}$ case using a resolution parameter $y_{{\rm cut}}$%
. This is a sensible way to define an {\em exclusive} jet cross section in
analogy to the $e^{+}e^{-}$ case but now including the beam jets. For
instance, one might measure in this way a cross section $\sigma _3(d_{{\rm %
cut}},y_{{\rm cut}})$ to produce exactly three high $E_T$ jets plus the beam
jets. However, we prefer to maintain a similarity with the previous cone
algorithm work in hadron-hadron collisions. The jet definition in the
preceding section is intended to define {\em inclusive} jet cross sections
in terms of the single angular resolution parameter $R$ (which plays a role
similar to $y_{{\rm cut}}$). For example, the two jet inclusive cross
section $d\sigma /dM_{JJ}$ defined in this way is a function of only the
jet-jet invariant mass $M_{JJ}$ and $R$. It would be an additional
complication if it also depended on the parameters $d_{{\rm cut}}$
and $y_{\rm cut}$.

We note, however, that an exclusive n-jet cross section can be defined using
the algorithm in Section 2. In this case, one needs a jet hardness parameter
to play the role of $d_{{\rm cut}}$. A convenient choice is to count only
jets with transverse energies above a cutoff $E_{T,{\rm cut}}$. Except for
the issue of variable $R$, this is essentially the $y_{{\rm cut}}=1$
scenario of Catani {et al.}

The function $d_{ij}$ given in eq.~(\ref{dij}) measures how ``near\-by'' the
pair of protojets $(i,j)$ is. As in other algorithms of the successive
combination type, the idea is to combine first the protojets that are
``nearest'', and thus have the smallest $d_{ij}$. There are, of course,
other possibilities for the function $d_{ij}$, the measure of ``nearness''.
For instance, one might use the invariant pair mass [for massless protojets $%
i$ and $j$],
\begin{equation}
d_{i,j}^M  =  M_{ij}^2
  =  2E_{T,i}E_{T,i}[\cosh (\eta _i-\eta _j)-\cos (\phi
_i-\phi _j)].
\end{equation}
For small $|\eta _i-\eta _j|$ and $|\phi _i-\phi _j|$ this is
\begin{equation}
d_{i,j}^M\approx E_{T,i}E_{T,j}[(\eta _i-\eta _j)^2+(\phi _i-\phi _j)^2].
\end{equation}
Such a choice for $d_{i,j}$ yields an algorithm analogous to the ``Jade''
version of the successive combination algorithm used in $e^{+}e^{-}$
annihilation. The corresponding algorithm with the factor $E_{T,i}E_{T,j}$
replaced by $\min (E_{T,i}^2,E_{T,j}^2)$, as in eq.~(\ref{dij}), is
analogous to the ``Durham'' algorithm \cite{Durham} in $e^{+}e^{-}$
annihilation as noted earlier. A discussion of the relative merits of these
choices in the context of $e^{+}e^{-}$ annihilation may be found in \cite
{BKSS}. The Durham algorithm has been also discussed in the context of
lepton-hadron collisions in \cite{catani}.

\section{Comparison with cone algorithm}

Now consider the algorithm advocated here. At any stage in the operation of
the algorithm, the two protojets $i$ and $j$ with the smallest value of $%
d_{ij}$ are merged if $d_{ij}$ of eq.~(\ref{dij}) is less than the smaller
of $E_{T,i}^2$ and $E_{T,j}^2$. That is, they are merged if
\begin{equation}
\label{merge}\sqrt{(\eta _i-\eta _j)^2+(\phi _i-\phi _j)^2}<R.
\end{equation}
Thus the issue of which protojets to merge first depends on transverse
energies and angles, but the issue of whether to merge two protojets or to
declare that they cannot be merged is solely a question of the angle between
them.

The merging condition in eq.~(\ref{merge}) makes it clear that the present
algorithm is in fact not so different from a cone algorithm. The latter is
typically defined\cite{snowmass} in terms of the particles $n$ whose momenta
$\overrightarrow{p_n}$ lie within a cone centered on the jet axis ($\eta
_J,\phi _J$) in pseudorapidity $\eta $ and azimuthal angle $\phi $,
\begin{equation}
\label{merge2}\sqrt{ (\eta _n-\eta _J)^2+(\phi _n-\phi_J)^2} <R \, .
\end{equation}
The jet angles ($\eta _J,\phi _J$) are the averages of the particles'
angles,
\begin{mathletters}
\label{snow}
\begin{eqnarray}
   \eta_J  & = & \sum_{n\in {\rm cone}} p_{T,n} \eta_n/E_{T,J} \,\, ,
\\
   \phi_J  & = & \sum_{n\in {\rm cone}} p_{T,n} \phi_n/E_{T,J} \,\, ,
\label{axis2}
\end{eqnarray}
\end{mathletters}
with
\begin{equation}
\label{ET2}E_{T,J}=\sum_{n\in {\rm cone}}p_{T,n} \,,
\end{equation}
analogous to eqs.~(\ref{axis1}) and (\ref{ET1}), respectively. This process
is iterated until the cone center matches the jet center
($\eta _J,\phi _J$) computed in eq.~(\ref{snow}).

The definition of the jet axis given in eqs.~(\ref{snow}) and (\ref{ET2})
is chosen to be simple when expressed in the natural variables
$(E_T,\eta,\phi)$. Of course, other choices could also yield infrared
safe jet definitions.  Thus this definition should be regarded as a
convenient convention.  This convention has been continued in the merging
conditions, eqs.~(\ref{ET1}) and (\ref{axis1}), of the successive combination
algorithm.

While the successive combination algorithm never assigns a particle to more
than one jet, this is not the case for the cone algorithm as defined so far.
It is possible for jet cones to overlap, so that one
particle is contained in more than one jet. This issue was discussed
in \cite{EKS} in the context of the order $\alpha _s^3$
perturbative calculation. At
this order it is possible for two one-parton jets to lie within the cone of
a two parton jet. In the calculation\cite{EKS2}, such jets
are ``merged.''  That is, the
two parton jet is kept and the one parton subjets are not considered as
being legitimate jets on their own. In a physical event, with many more
particles, the merging question is more serious, and a criterion for merging
must be part of the experimental algorithm\cite{CDFmerge}.

The difference between cone and successive combination algorithm jets is
apparent even in the simplest example of merging two partons (or hadrons) to
make a jet. In the cone algorithm \cite{snowmass}, two partons $i$ and $j$
are merged if each falls within an angular distance $R$ of the jet axis
defined by eq.~(\ref{axis2}). The parton with the smaller $E_T$, call it
$i$, is farther
from the jet axis. In the cone algorithm, the limit on angular
separation between this parton and the jet $J$ (including the partons $i$
and $j$) has the simple form
\begin{equation}
\label{coneangle}\sqrt{(\eta _i-\eta _J)^2+(\phi _i-\phi _J)^2}<R\,.
\end{equation}
The corresponding relation in terms of the angular separation of the two
partons is
\begin{equation}
\label{anglecone2}\sqrt{(\eta _i-\eta _j)^2+(\phi _i-\phi _j)^2}<{\frac{{%
E_{T,J}}}{{E_{T,J}-E_{T,i}}}}\,R\,.
\end{equation}
The limit on the right hand side lies in the range $R<%
\{E_{T,J}/[E_{T,J}-E_{T,i}]\}R\leq 2R$ since $0<E_{T,i}\leq E_{T,J}/2$. Thus
configurations are possible with two equally energetic partons located near
opposite edges of the cone, and nothing in the center of the cone. These are
precisely the configurations where the merging issue arises in order $\alpha
_s^3$ perturbation theory.

In the successive combination algorithm, it is the separation between the
two partons (or protojets) that has the simple limit of eq.~(\ref{merge}).
It is this angle between $i$ and $j$, not that between $i$ and the final jet
direction, that controls the question of merging. The corresponding relation
for the angular separation between parton $i$ (the lower $E_T$ parton), and
the jet $J$ in order for merging to occur is
\begin{equation}
\label{succangle}\sqrt{(\eta _i-\eta _J)^2+(\phi _i-\phi _J)^2}<{\frac{{%
E_{T,J}-E_{T,i}}}{{E_{T,J}}}}\,R\,,
\end{equation}
where the right hand side is in the range $R/2\leq [{{E_{T,J}-E_{T,i}}]/{%
E_{T,J}}}<R$. Thus the lower-$E_T$ parton can be far from the jet axis, up
to a maximum separation $R$, while the higher $E_T$ parton must be closer
to the jet axis.

Because of the difference between eq.~(\ref{coneangle}) and eq. (\ref
{succangle}), the average distribution of transverse energy within jets
depends on which algorithm one uses. With the successive combination
algorithm, there is less transverse energy near the edge of the allowed
angular region than there is with the cone algorithm. This is illustrated in
Fig.~1. The quantity plotted is the order $\alpha _s^3$ perturbative result
for the average transverse energy fraction as a function of distance from
the center of the jet, for jets with $E_T=100$ GeV at $\sqrt{s}=1800$ GeV
for $R=1$. The histogram represents the $E_T$ fraction in angular annuli, $r$
to $r+0.1$, where $r$ is the distance from the jet center, $r=\sqrt{(\eta
-\eta _J)^2+(\phi -\phi _J)^2}$. Thus the sum over all bins, $r<R$ including
$0<r<0.1$, should equal unity. Note also that the calculation yields energy
outside of the cone, $r>R$. This arises in the perturbative calculation from
configurations with only one parton in the cone (the jet) but with another
parton nearby.

The cone algorithm, in principle, requires one to find all possible
solutions of the cone matching conditions, eqs.~(\ref{merge2}), (\ref{snow})
and (\ref{ET2}), before beginning
the merging algorithm. In experimental
applications\cite{CDFmerge}, however, one may begin the search for
valid jet cones using ``initiator'' calorimeter cells, {i.e.}, cells
with $E_T$ above some threshold value. In this case, possible jets that
consist of two widely separated subjets may not be recognized because there
is no initiator cell between the subjets. Thus, the two widely separated
jets may not be merged. The theoretical study \cite{EKS} suggests that the
jet finding algorithm used by the CDF group is likely not to merge two
subjets when their separation is greater than a value $R_{sep}$, which
appears to be about $1.3R$ in practice as is consistent with CDF studies%
\cite{CDFmerge}. In this language, the successive
combination algorithm, when applied to two parton jets, corresponds to the
limiting case $R_{sep}=R$.

At higher orders in perturbation theory or for more realistic jets
containing many particles, it is eq.~(\ref{coneangle}) that survives in the
cone algorithm case and eq.~(\ref{merge}) for the successive combination
algorithm. Thus cone jets always have
well defined, smooth boundaries, although the amount
of $E_T$ near the edge will depend in practice on how the issue of the
merging of overlapping jets is handled. For
the successive combination algorithm there is no ``merging
question.'' It is automatically dealt with by the algorithm as all particles
are assigned to a unique jet. Furthermore, only small $E_T$ particles can
be as far as $R$ from the jet axis. However, there is a price for this
simplicity. Particles with very small $E_T$ can be very far from the jet
axis ($>R$). Thus jet boundaries can be complicated. Here, the issue is
intertwined jets rather than overlapping jets.

Does it follow that the successive combination algorithm is better in some
sense than the cone algorithm, or is it just different? The criteria that we
will examine are 1) estimated size of higher order perturbative corrections
to a jet cross section calculated using fixed order perturbation theory, 2)
estimated size of corrections to the calculated cross section arising from
``power suppressed'' or ``hadronization'' effects, and 3) simplicity and
definiteness of the algorithm.

We turn first to the estimated size of perturbative corrections. We consider
as an example the one jet inclusive cross section $d\sigma /dE_Td\eta $ at $
\sqrt{s}=1800$ GeV, averaged over the rapidity range $0.1<|\eta |<0.7$ used
by the CDF Collaboration where the comparison of theory with data has been
very successful\cite{EKS2,CDF}. Again we calculate this cross section at order
$\alpha _s^3$ according to both jet definitions, using the methods described
in \cite{EKS} and \cite{EKS2}. The question is then,
how big are the order $\alpha _s^4$
corrections? Of course, we can only make estimates. One way to do this is to
examine the dependence of the calculated cross section on the
renormalization and factorization scale $\mu $. If we could calculate to
order $\alpha _s^4$, then the renormalization group guarantees that some of
the $\alpha _s^4$ terms would cancel most of the $\mu $ dependence that
occurs in the $\alpha _s^3$ cross section (so that the derivative of the
cross section with respect to $\mu $ would then be of order $\alpha _s^5$).
Thus the $\alpha _s^4$ contributions to the cross section are likely to be
at least as large as the difference between the cross section calculated
with a ``best guess'' for $\mu $, which we take to be $\mu =E_T/2$, and the
cross section calculated with the alternative choices $\mu =E_T/4$ or $\mu
=E_T$. In Fig.~2a, we show the cross section at $E_T=100$ GeV as a function
of the clustering parameter $R$ calculated according to both algorithms for
these three values of $\mu $.

This rather busy graph becomes remarkably simple, as in Fig.~2b, if we
rescale the parameter $R$ in the two cases. Let the clustering parameter in
the successive combination algorithm, eq.~(\ref{dij}), be called $R_{{\rm %
comb}}$ for the present purposes and plot the cross section versus
\begin{equation}
R^{\prime }=R_{{\rm comb}}.
\end{equation}
For comparison, we label the clustering parameter in the case of the cone
algorithm, eq.~(\ref{merge2}), as $R_{{\rm cone}}$ and in Fig.~2b we display
the cone results as a function of a scaled parameter
\begin{equation}
R^{\prime }=1.35\times R_{{\rm cone}}.
\end{equation}
We see from the graph that, for each value of $\mu $, the calculated cross
sections are almost identical as long as we identify $R_{{\rm comb}}\approx
1.35\times R_{{\rm cone}}$. Thus, for instance, a jet cross section
calculated or measured with the cone algorithm using the standard value
$R_{{\rm cone}}=0.7$ should be compared to a jet cross section with the
successive combination algorithm using $R_{{\rm comb}}=0.945$ $\approx 1.0$.

In Fig.~3 we plot the
order $\alpha_s^3$
inclusive jet cross section as defined by the successive combination
algorithm with $R_{\rm comb} = 1.0$ versus the jet $E_T$ at
$\sqrt{s} = 1800$ GeV taking $\mu = E_T/2$.  The corresponding cross section
calculated using the cone algorithm with $R_{\rm cone} = 0.7$ is nearly
identical (see, for example, Fig.~1 of \cite{CDF}), and is not shown as it
would not be distinguishable.  In the range $E_T/4 < \mu < E_T$,
10 GeV $< E_T <$ 500 GeV, we find that the two algorithms give order
$\alpha_s^3$ theoretical cross sections that agree to within 10\%.

We conclude from this comparison that the successive combination algorithm
is neither better or worse than the cone algorithm, at least as judged
according to this $\mu $-dependence standard. We also note that the
argument\cite{EKS} that $R_{{\rm cone}}\approx 0.7$ is a
particularly stable and
thus sensible regime for comparing fixed order perturbation theory with
experiment is now translated into $R_{{\rm comb}}\approx 1.0$. This is just
the regime studied by Catani {et al.} \cite{Webber}.

We now examine the question of higher order corrections from a point of view
that relates directly to the reasoning behind the successive combination
algorithms: the idea of putting parton showers back together to reconstruct
the parent parton. We suggest that jet cross sections defined with the cone
algorithm may have larger higher order perturbative corrections than do jet
cross sections defined with the successive combination algorithm because of
what may be called ``edge of the cone'' effects. Consider the application of
the cone algorithm to two partons, 1 and 2, with roughly equal transverse
energies and an angular separation of approximately $2R_{{\rm cone}}$, the
troublesome configuration mentioned earlier. These two partons are near the
edge of the region in which they can form an allowable cone jet. Suppose
that parton 2 splits into two partons, 2a and 2b, that each have substantial
transverse energy. If the angle separating partons 2a and 2b is
infinitesimal, then they will both fit into the jet cone. The resulting jet
(1,2a,2b) will have the same direction and transverse energy as the jet
(1,2) that one obtains if the splitting did not occur. However, if the angle
separating partons 2a and 2b is small but not infinitesimal, it can very
well be that the three partons (1,2a,2b) cannot fit into a cone to form a
single jet. Since the matrix element for such a parton splitting is large
(although not infinite), one may worry that there will be corresponding
large order $\alpha _s^4$ corrections to jet cross sections calculated with
the cone algorithm. What is the situation with the successive combination
algorithm? Here we should consider partons 1 and 2 separated by an angle of
approximately $R_{{\rm comb}}$, near the edge of the region in which they
can form an allowable jet.  If parton 2
splits into partons 2a and 2b with a small angular separation, then the
successive combination algorithm will combine them together into a protojet $%
2^{\prime }$ that approximates parton 2. As long as the angle separating
partons 2a and 2b is not too large, the protojet $2^{\prime }$ will have jet
parameters close enough to those of parton 2 that the algorithm will then
combine $2^{\prime }$ with parton 1. On the basis of this argument, we
expect that order $\alpha _s^4$ corrections to jet cross sections calculated
with the successive combination algorithm may be smaller than with the cone
algorithm. For similar reasons, we expect also that jet cross sections
calculated with the successive combination algorithm will also exhibit
smaller corrections attributable to the final combination of partons into
hadrons. Unfortunately, from the order $\alpha _s^3$ perturbative
calculation we cannot determine the magnitude of these edge of the cone
effects.
\footnote{This is clearly a subject for Monte Carlo study as in \cite{Webber}.
Unfortunately the cone algorithm used in that analysis
is quite different from that described here.  In particular, the cone
algorithm in \cite{Webber}
is not infrared safe so that the results presented are difficult to
interpret.}

Finally, we comment on the relative merits of the two algorithms from the
point of view of simplicity and definiteness. Here the cone algorithm
appears at first to have the advantage. With the cone algorithm, a jet
consists simply of all the particles whose momentum vectors fit into an
apparently well defined and regular cone centered on the jet axis. This is a
simple and appealing idea. The successive combination algorithm, in
contrast, takes more effort to define and does not yield regular jet shapes
in the $\eta -\phi $ plane. However, in the application of the cone
algorithm, one quickly discovers that ambiguous cases with overlapping jet
cones arise. What do you do when two or more jets contain
particles in common? The algorithm must be expanded to cover these cases.
Unfortunately, there are many ways to proceed, and none of them is
particularly simple. Thus the cone algorithm actually consists of a simple
part that works beautifully for joining two partons into a jet and a
complicated part that one is forced to use in order to deal with real-world
multi-hadron events. With the successive combination algorithm, the
recursive steps require some thought to define, but once they are set, the
whole algorithm is complete.

\section{Summary}

We have discussed a jet definition for inclusive jet measurements in hadron
collisions that makes use of ideas previously applied to jet definitions for
$e^+ e^-$ collisions. The
algorithm identifies a jet by successively combining ``nearby'' pairs of
particles or
protojets.  The concept of ``nearby'' is measured in $(E_T,\eta,\phi)$
space and involves a limit $R_{\rm comb}$ on angular separations
that is very similar to the usual cone
algorithm parameter $R_{\rm cone}$.  We find that, calculated perturbatively,
the inclusive jet cross section that results from the new algorithm with
parameter $R_{\rm comb}$ is
essentially identical to the cone algorithm result with $R_{\rm cone} =
R_{\rm comb}/1.35$.  While the final geometry of a jet defined by the
successive combination algorithm is likely to be more complex than that from
the cone algorithm, the former definition has the advantage that there is no
problem with overlapping jets as there
is in the cone case.  We have also presented a qualitative argument
that, due to ``edge of the cone'' effects, cross sections calculated with
the successive combination algorithm are likely to exhibit smaller higher
order and hadronization
corrections.  Only further detailed experimental and
theoretical studies can demonstrate whether the successive combination
type algorithm has quantitative
advantages over other algorithms.

\acknowledgments

We thank the members of the CDF Collaboration's QCD group, and in particular
J. Huth and N. Wainer, for discussions concerning the CDF jet measurements.
We also express our appreciation to S. Catani, M. Seymour and B.R. Webber
for discussing their analysis with us. Finally we thank Z. Kunszt and
D. Kosower for helpful discussions.
This work was supported in part by
U.S. Department of Energy grants DE-FG06-91ER-40614 and DE-FG06-85ER-40224
and by the CERN TH Div, whom SDE thanks for their hospitality.

\begin{figure}
\caption{
Fraction of jet $E_T$ in angular annuli $r$ to $r+0.1$
comparing the cone algorithm
with the the successive combination case. In both cases the
jet has $R = 1.0$, $E_T =100$ GeV, $\protect\sqrt{s} = 1800$ GeV,
$0.1 < |\eta_J| < 0.7$ with renormalization/factorization scale
$\mu = E_T/2$ and the structure functions of HMRS(B)\protect\cite{HMRS}.
}
\label{annulus}
\end{figure}

\begin{figure}
\caption{
Order $\alpha_s^3$ inclusive jet cross
section for $E_T=100$ GeV, $\protect\sqrt{s}=1800$ GeV,
averaged over $\eta_J$ in the range
$0.1<|\eta_J|<0.7$ with the structure functions of
HMRS(B)\protect\cite{HMRS}
for the  two algorithms specified in the text.
The curves for the cone algorithm are: $\mu = E_T$ (solid), $\mu = E_T/2$
(dot-dash), $\mu = E_T/4$ (dot-dot-dot-dash); for the successive combination
algorithm: $\mu = E_T$ (long dash), $\mu = E_T/2$ (medium dash),
$\mu = E_T/4$ (short dash).
a) Standard case plotted versus $R = R_{\rm cone} = R_{\rm comb}$.
b) Same as a) except that the cone algorithm is plotted
versus $R^\prime= 1.35 R_{\rm cone}$ while the successive combination
case has $R^\prime=R_{\rm comb}$.
}
\label{Rfig}
\end{figure}

\begin{figure}
\caption{
Order $\alpha_s^3$
inclusive jet cross section as defined by the successive combination
algorithm with $R_{\rm comb} = 1.0$ versus the jet $E_T$ for
$\protect\sqrt{s} = 1800$ GeV,
$\mu = E_T/2$, averaged over $\eta_J$ in the range
$0.1<|\eta_J|<0.7$ with the structure functions of
HMRS(B)\protect\cite{HMRS}.
}
\label{sigma}
\end{figure}

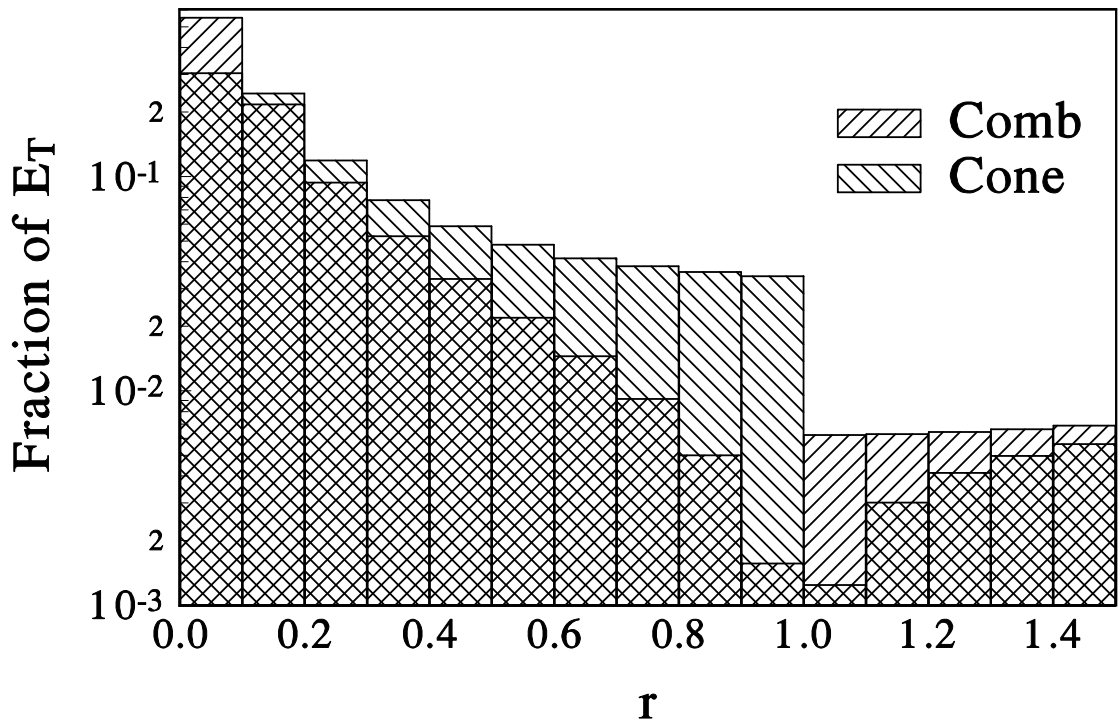
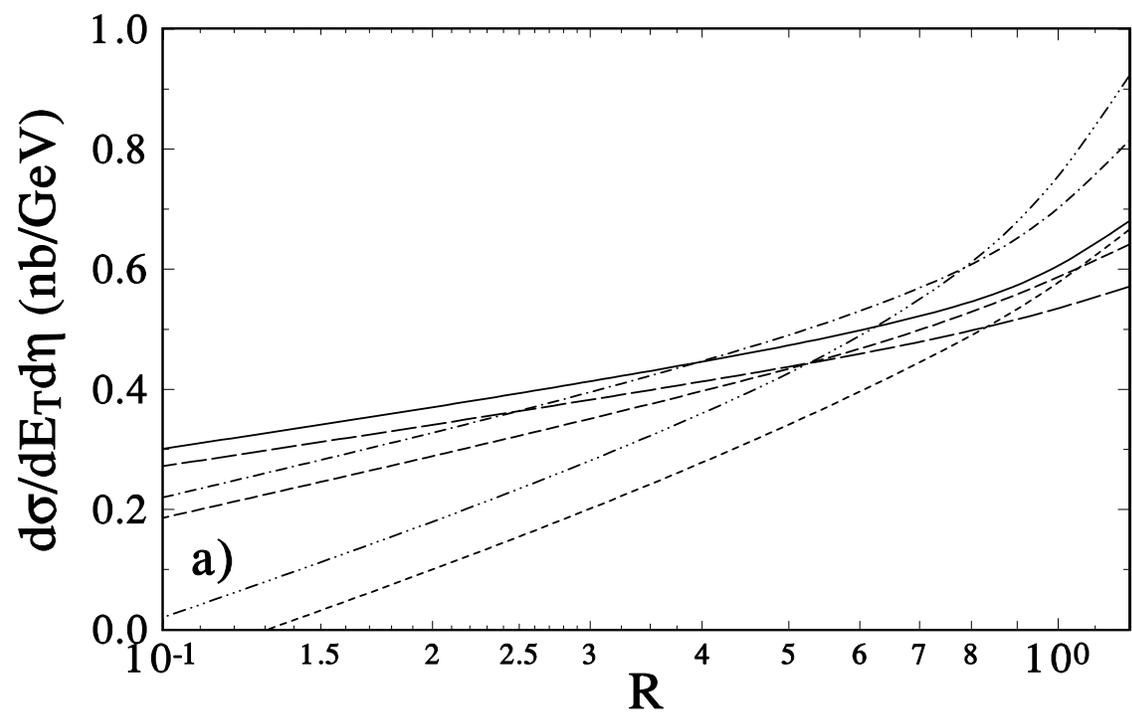
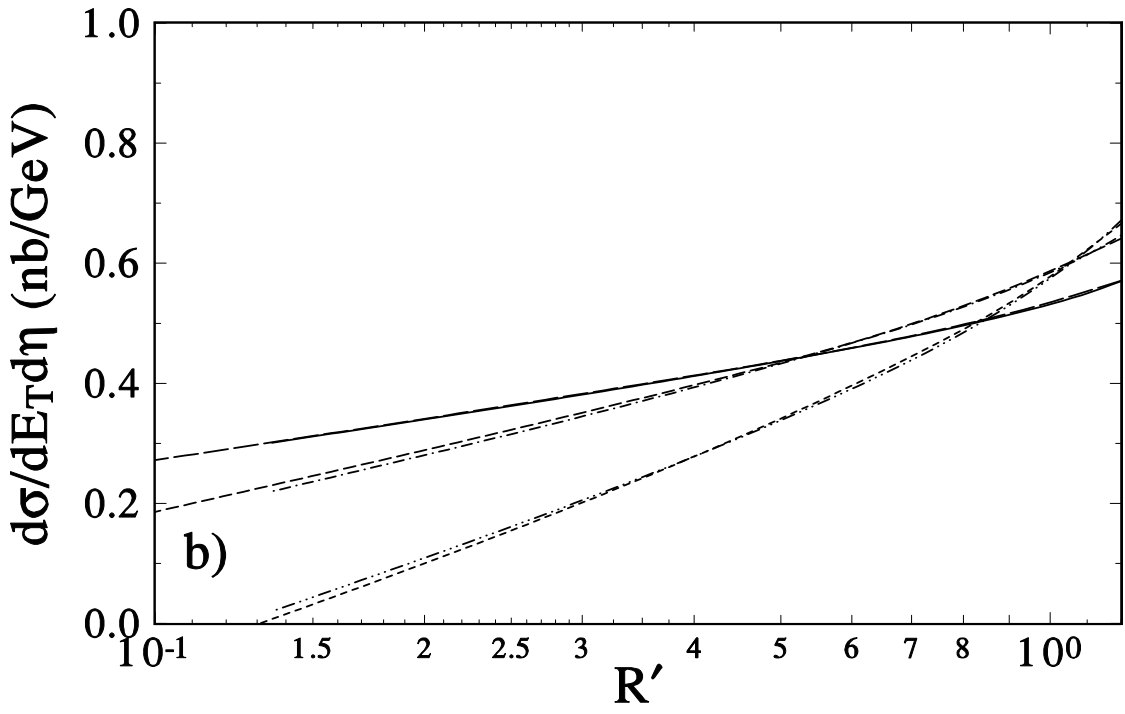
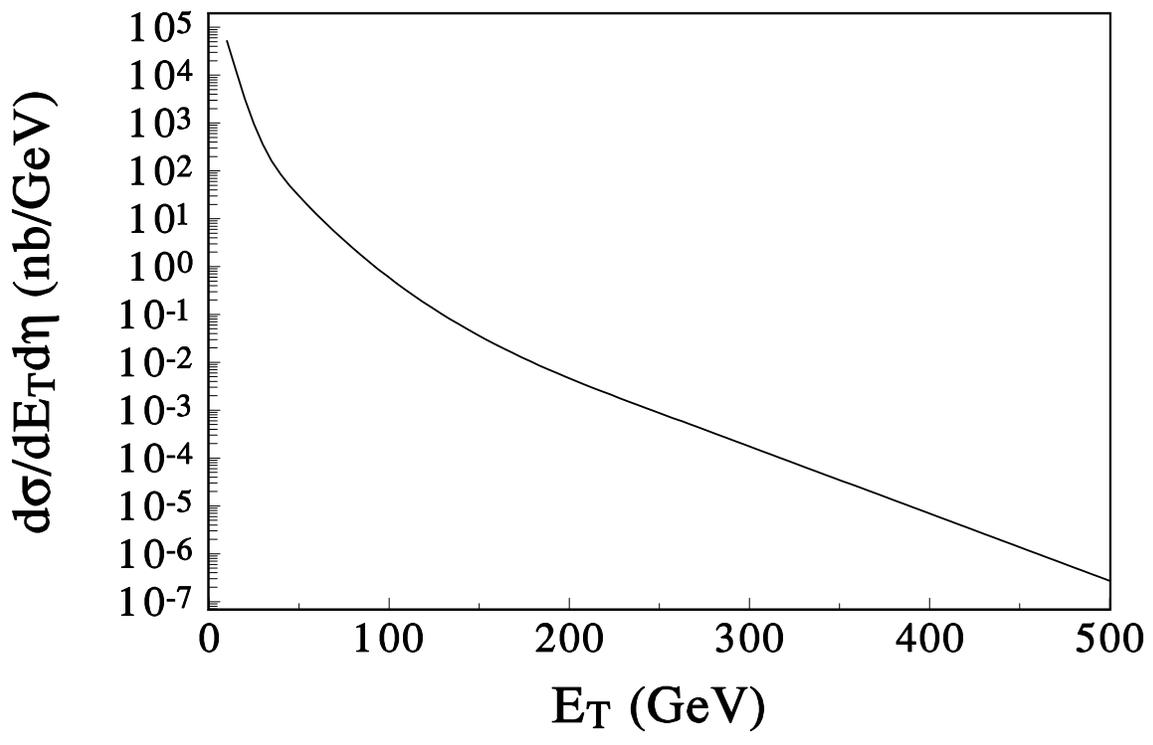
\end{document}